\documentclass[sigconf]{acmart}

\usepackage{booktabs} 
\usepackage{algorithmic}
\usepackage{algorithm}
\usepackage{flushend}

\graphicspath{ {images/} }

\DeclareMathOperator*{\argmax}{arg\,max}

\renewcommand\footnotetextcopyrightpermission[1]{} 
\acmConference[KDD'2018]{KDD conference}{August 2018}{London}
\acmYear{1997}

\author{Shreya Singh}
\affiliation{%
  \institution{Myntra Designs, India.}
}
\email{shreya.singh1@myntra.com}

\author{G Mohammed Abdulla}
\affiliation{%
  \institution{Myntra Designs, India.}
}
\email{ghani.abdulla@myntra.com}

\author{Sumit Borar}
\affiliation{%
  \institution{Myntra Designs, India.}
}
\email{sumit.borar@myntra.com}

\author{Sagar Arora}
\affiliation{%
  \institution{Myntra Designs, India.}
}
\email{sagar.arora@myntra.com}

\begin{document}
\title{Footwear Size Recommendation System}

\begin{abstract}
While shopping for fashion products, customers usually prefer to try-out products to examine fit, material, overall look and feel. Due to lack of try out options during online shopping, it becomes pivotal to provide customers with as much of this information as possible to enhance their shopping experience. Also it becomes essential to provide same experience for new customers. Our work here focuses on providing a production ready size recommendation system for shoes and address the challenge of providing recommendation for users with no previous purchases on the platform. In our work, we present a probabilistic approach based on user co-purchase data facilitated by generating a brand-brand relationship graph.  Specifically we address two challenges that are commonly faced while implementing such solution. 1. Sparse signals for less popular or new products in the system 2. Extending the solution for new users. Further we compare and contrast this approach with our previous work \cite{abdullasize} and show significant improvement both in recommendation precision and coverage. 

\end{abstract}

%
%


\keywords{Collaborative Filtering, Probabilistic Graphical model, Co-occurence graph, Word2vec, Brand-Brand similarity}

\maketitle

\section{Introduction}

Online shopping for fashion products is growing at astounding pace. In order to entice users on the platform and ease their shopping experience, typically e-commerce platforms provide free return shipping. Compared to offline shopping, on online platforms users typically lack the option of trying out products before purchase. This limits users
from exploring several important aspects of fashion shopping like how it fits, looks or feels ? While purchasing footwear, size or fit is an extremely critical part of the selection process and often the deciding factor. Lack of good fit results in user returning the product ensuing bad shopping experience and increased operational cost for the platform \cite{internetRetailer}.  

Further, on an e-commerce platform users are presented with hundreds of brands and typically these brands do not follow similar sizing guideline \cite{hsu2009data}. For example a sports shoe from Nike of UK size 9 would have 27.5cm length while an Adidas shoe of same UK size would have length of 27.0cm. Shoe length is just one attribute of the product and various other attributes like arch type, shoe width, shoe type play a key role in finding the right fit. So choosing the right size is an unpleasant experience due to lack of standardization in fashion industry. Hence, a system to recommend the right size for users is critical in easing user's online shopping journey and providing better overall experience. 

 Myntra is the largest Fashion E-commerce company in India with over 3 million products hosted on its platform. Everyday millions of users come to our platform and make purchases. So we want to leverage the rich purchase history of the users' purchases to make recommendations.  The problem of size recommendation is further complicated as lot of important sizing related attributes are either missing or inaccurate. Typically for footwear, most vendors only provide length of the footwear and 80\% of the vendors do not provide any information on shoe width, shoe arch type etc.  

There have been various attempts in industry to solve the size recommendation problem \cite{hu2015collaborative,adomavicius2005toward,aroradecoding,bracher2016fashion}. Amazon proposed a Size Recommender System \cite{sembium2017recommending} , which relies on learning a one dimension latent feature for every product and user. However, that method has the disadvantage that it cannot be used with sparse data as latent features have to be learnt for every product. 
 In our previous work \cite{abdullasize}, we had presented a size recommendation system for apparel by leveraging the purchase data and learning latent features from products using Word2vec \cite{mikolov2013efficient}.  In this setting, products purchased by the users are considered as words and all the products purchased by one user is taken to be a document. The latent embeddings are learnt for each product using Word2vec. Then latent embeddings of products purchased by a user are combined to form a user embedding. A simple inner product between a user and product would give a product fit score for the given user and product. Although this method is very effective in making apparel size recommendation, it cannot be used to make footwear recommendation.  The main reason being; users do not make frequent footwear purchases as they would buy apparels. A strawman approach of predicting same size across brands would also not work due to lack of standardization in sizes. Sparsity is hereafter defined as $1 - \ edge density$ where edge density of graph is defined as ratio of number of edges in graph to the corresponding complete graph.

The sparsity in the brand-brand co-purchase graph is $0.64$ which means most of the brands are either not co-purchased or are co-purchased rarely. Our approach Weighted Brand Similarity Recommendation (WBSR) addresses this challenge by creating a brand-brand relationship graph using our clickstream data (clicks, carts, orders etc.) We augment the co-purchase graph with brand-brand similarity scores for making the recommendations. The approach is discussed in detail in Section \ref{wbsr}. We compare this approach with another Skip-gram based Word2vec routine which predicts the best product size based on the cosine similarity of the latent vector representations of footwear SKUs. A SKU is a style-size combination index which identifies every product individually. Being completely based on purchase signals and product history for every brand, this approach considerably suffers from sparsity. On contrary, WBSR which works on a higher abstraction level of brands helps to overcome sparsity. 

\begin{figure}
\includegraphics[height=3.2in, width=3.5in]{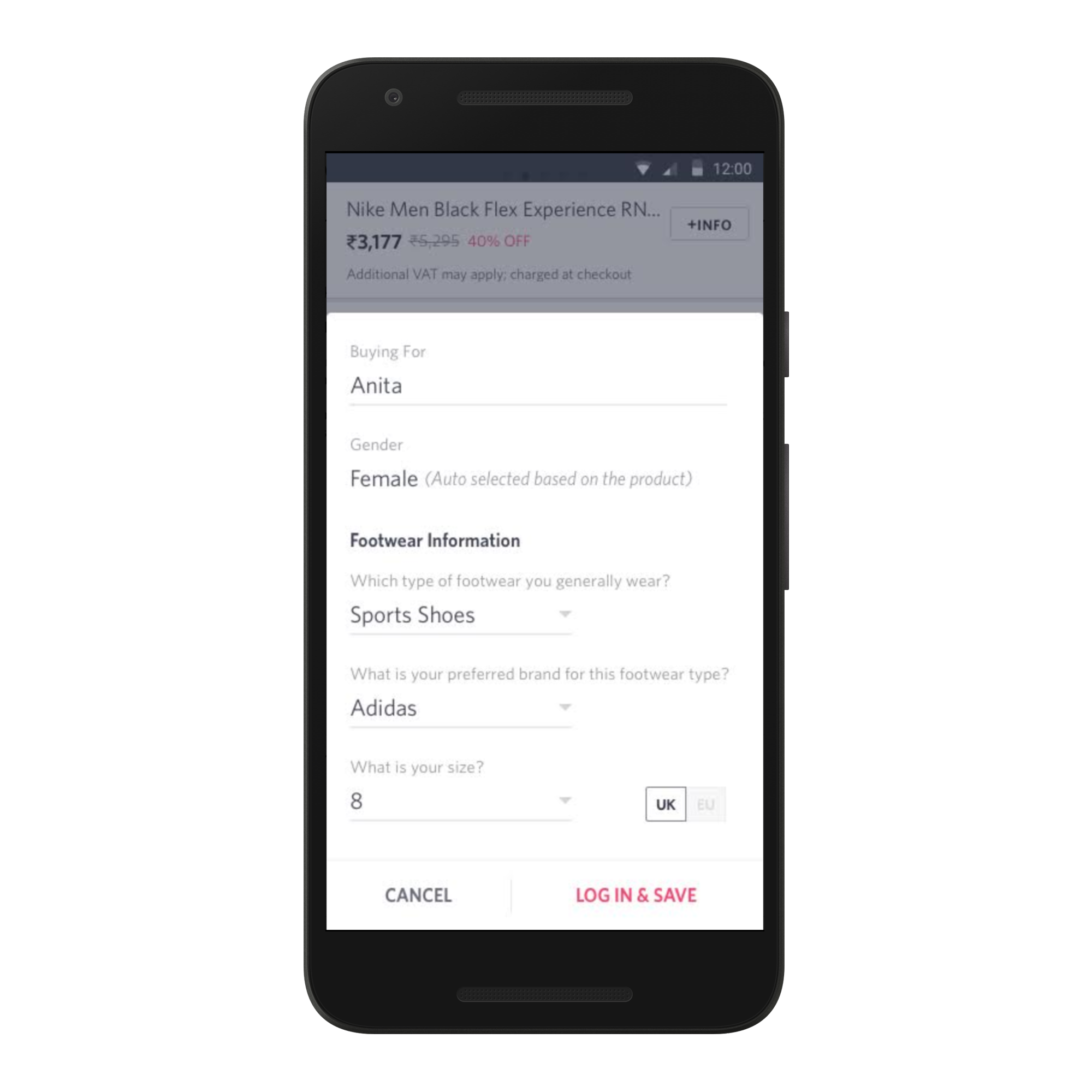}
\caption{Footwear Preference Questionnaire}
\label{fig:phone}
\end{figure}

Another challenge we address is the cold start problem where users have not made a single purchase before. We could start making size recommendations for them by asking simple questions like "which brand do they prefer in certain category?" and "what size do they typically buy?" when they come on our platform for the first time. This can be seen in Figure \ref{fig:phone}.  We show that answers to these simple questions are enough to make effective size recommendations, which in turn reduce returns on the platform. 

The rest of the paper is organized as follows. Section 2 contains the methodology of the Skip-gram based recommender system and WBSR. We explain the system diagram in detail along with necessary examples in this section. This is followed by Results and Analysis section which consists the experimental results and Dataset statistics. We have also covered the graphical and brand-brand similarity statistics in this section. 

\section{Methodology}
The intuition behind solving the footwear size recommendation problem through a probabilistic graphical-based approach lies very much on the fundamental properties of a graph. The idea is to represent the nodes of the graph as products and the edges as how many times the products are bought together. Modeling the size recommendation problem as a co-purchase graph not only captures the latest buying trends across brands but also helps to learn the size inconsistency across them. In this section, we would be discussing in detail two different graphical methods we have used and how each one caters to solve the footwear recommendation problem. We would also be discussing the shortcomings observed and solutions to them. For providing good recommendations, we generate the graphical models at a gender-article type level (e.g. Men Sports Shoes, Men Casual Shoes). For the approaches mentioned, we would be splitting data at a gender-article type level and would be using the UK unified size across all the brands.

\subsection{Skip-gram based size recommendation}

Our first approach to solve footwear recommendation is based on representing the user co-purchase history in the size and fit space. Consider the user purchase data as a sparse matrix $W$ where each row is a user and the columns represent the SKUs bought by the user. The aim is to learn a joint probability function for the SKUs bought together by the user through Word2Vec which would generate latent vectors for the SKUs. In \cite{abdullasize}, the authors talk about this approach wherein they represent each SKU as a combination of size and fit attributes. For example, a SKU with gender  "Men", article type "Shirts", brand "Allen Solly", fit "Slim", usage attribute "Casual" and size "M" would be encoded as "Men\_Shirts\_Allen\_Solly\_Casual\_M". The intuition behind training a Word2vec model over the matrix $W$ is that users generally buy products of same size, hence the Word2Vec model would project out clusters of same-sized SKUs in the size and fit space.

We employ a similar approach for our problem, with some changes. In the method described above, the SKUs are represented at a much granular level, with many features taken into account (usage attribute, occasion, fit). However, this can further infuse sparsity into the size and fit space especially in the case of footwear as people buy much lesser footwear than apparel. Hence, we encode the footwear SKUs in a similar fashion, but only include the brand and size information. For our use case, we would represent a SKU with gender "Men" article type "Casual Shoes" brand "Vans" and UK size "8" as "Men\_Casual\_Shoes\_Vans\_8". The SKUs can be considered as words and all the SKUs purchased by a user can be considered as a document. With these words and documents as inputs we train the Word2Vec network.

The Word2Vec network consists of the matrix $W$ where $W_{i}$ represents all the orders of a user i sorted by date which can be represented by the sequence ${W_{i1}, W_{i2}, W_{i3},..W_{in}}$, here each $W_{ij} \in C$ where $C$ is the entire platform catalog for a particular article type and $W_{ij}$ is the word representation of the SKU. The objective of the skip gram model is to maximize the log probability.

\begin{equation}
  \frac{1}{mn}\sum_{i=1}^{m} \sum_{j=1}^{n} \sum_{-q\leq k\leq q,k \neq 0} \log p(w_{i,j+k}\mid w_{i,j})
\end{equation}

where $q$ is the hyperparameter denoting length of the purchase window. Larger $q$ results in SKUs spanning over wide range of purchases to be considered as having same size and fit. The formulation of $p(w_{j}\mid w_{j+k})$ is given using Softmax function:

\begin{equation}
    p(w_j \mid w_{j+k}) = \frac{e^{u_{j},v_{j+k}}}{\sum_{k\in W}e^{u_k,v_k}}
\end{equation}

where $u$ and $v$ are the input and output one hot encoded vector representation of $w$. After we have trained the neural network, we compute the activation of the hidden layer for each SKU $c_p \in C$  and form a latent feature vector representation $f_p$.

Post training, the activation layer is scraped of and output of the hidden layer is used as the latent feature vector representation for each SKU. For making a footwear size recommendation, we make use of the explicit shoe preference questionnaire filled by the user,i.e., if the user has filled the shoe preference questionnaire as article type generally worn : Men Casual Shoes, preferred brand : Roadster , typical size worn : UK8, we would encode this information as a word and fetch the latent vector representation of this word. Let this representation be $U$. In real time, the latent feature vector representations (Word2vec vector) for each SKU or size for the Product page the user is currently on is fetched. Let this be ${V_{i}}$ where $i \in S$ where $S$ is size set of the product. We calculate the recommended product-size $i$ by using the following cosine similarity equation:

\begin{equation}
       i =  \argmax{(U . {V_{i}}) \ where \ i \in S } 
\end{equation}

The cosine similarity will measure the cosine of angle between two non-zero vectors which belong to an inner product space. Here, when we trained the Word2Vec neural network with the orders data, similar-sized SKUs formed clusters together in size and fit space which was represented through their latent feature vector representation. Hence, the non-uniformity of sizes among the brands is also taken care of as the users' order data spans across different brands. This means that the Word2vec network would learn the links between different size labels of different brands which essentially are of the same shoe measurement. This said, the latent feature vector representation of size $i$ which has the maximum cosine similarity value with $U$ is the recommended footwear size.  

\subsection{Weighted Brand Similarity Recommendation (WBSR)} \label{wbsr}

\begin{figure}
\includegraphics[height=2.0in, width=3.5in]{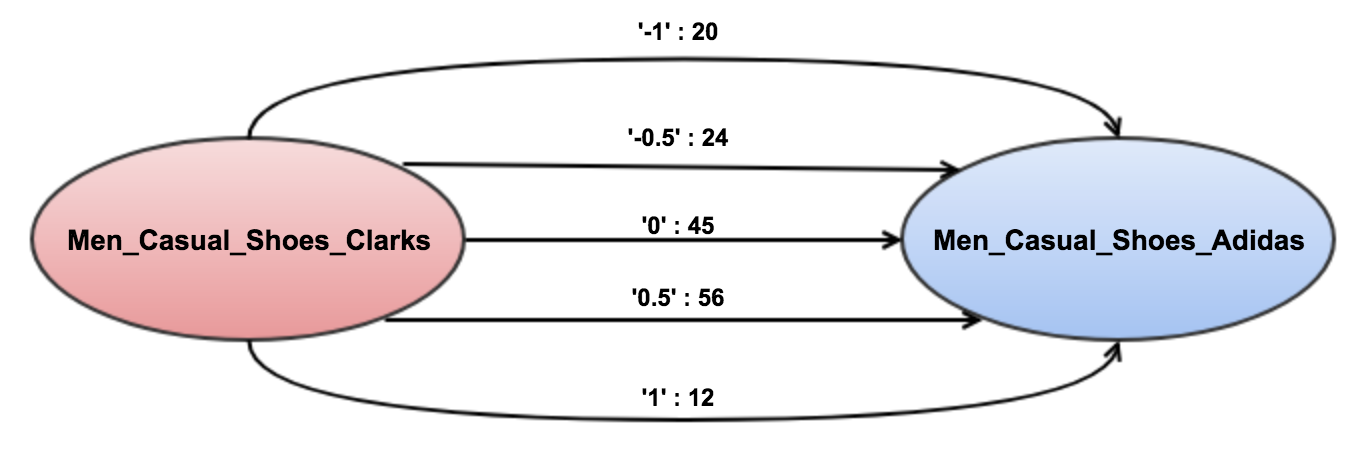}
\caption{Size Recommendation Graph}
\label{fig:size-graph}
\end{figure}

\begin{figure*}
    \centering
    \includegraphics[width=\textwidth]{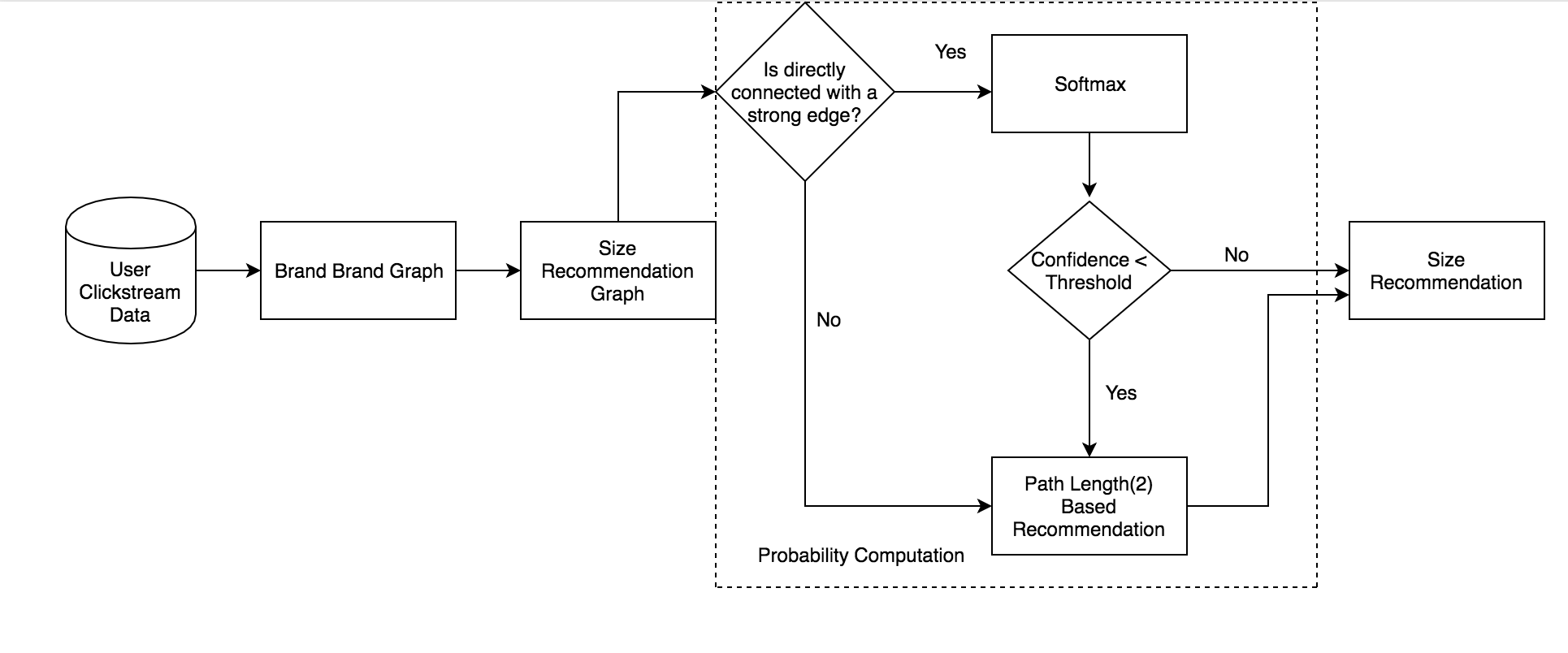}
    \caption{System Architecture}
    \label{fig:arch}
\end{figure*}

In this approach, we would be primarily building two graphs: Brand Brand Similarity Graph and the Size Recommendation Graph. As the name suggests the Brand-Brand similarity graph would give us insights about how similar two brands are; this information is then leveraged to build the Size Recommendation Graph.
Our brand-brand graph basis purchase interaction has an average sparsity of 0.8 for the approach mentioned in the above section. Most of the brands have either never been co-purchased or have been co-purchased rarely. 
This approach exploits the fact that purchase interactions, when augmented with other browse signals like (clicks, add to carts etc) can help to solve for sparsity. First we carve out a brand-brand graph basis the click stream data and thereafter use that to probabilistically model size recommendations. In this approach, we would be using two tune-able hyperparameters $\alpha$ and $\lambda$ which signify the total edge weight strength and softmax threshold respectively. The entire flow is shown in Figure \ref{fig:arch}

\subsubsection{Brand Brand similarity Graph}

As in \cite{aroradeciphering}, we use the following terminology hereafter.
\\
\\Let \(\textbf{U}\) be the set of all users, \(\textbf{B}\) be the set of all brands. Further, let \(m\) be the number of users and \(n\) be the number of brands.
\\
\\Let \(\textbf{E}_{ub} = \{E_c, E_b, E_w, E_p\}\) be the set of all possible events (user-brand interactions) where \(E_c = \) click event, \(E_b = \) add to bag event, \(E_w = \) add to wish list event and \(E_p = \) purchase event. It is worth mentioning that we just consider the event of highest priority for a given user and brand with priorities defined as:
\begin{equation}
E_c <  E_b < E_w <  E_p
\end{equation}
\\
We also define the importance score\footnote{The importance scores are computed using past 1 month data of the platform.} of an event \(e\) as 
\begin{equation} \label{eq:imp}
     w_e= 
\begin{cases}
    1,& \text{if } e = E_c\\
    \\\frac{\sum E_c} {\sum E_b},& \text{if } e = E_b\\
    \\\frac{\sum E_c} {\sum E_w},& \text{if } e = E_w\\
    \\\frac{\sum E_c} {\sum E_p},& \text{if } e = E_p
\end{cases}
\end{equation}

We construct the undirected weighted graph \(G_b = (\textbf{V},\textbf{E})\) with \(\textbf{V} = \textbf{B}\) and \textbf{E} be the edges denoting the similarity between two brands. Considering $e_{ij}$ as the interaction event of user \(U_i\) with brand \(B_j\), \(U_i\) is represented by a vector \(\textbf{u}_i\) of length \(n\) such that
\begin{equation}
     u_{ij}= 
\begin{cases}
    w_e,& \text{if } e_{ij} \ \in\  \textbf{E}_{ub}\\
    \\0,& otherwise
\end{cases}
\end{equation}
\\
We create a sparse user-brand matrix of size \(m \times n\) where each row represents a user and each column represents a brand. On this matrix, we apply non-negative matrix factorization \cite{lee2001algorithms} to learn the latent d-dimensional embeddings for each user (\(\textbf{v}_u\)) and brand (\(\textbf{v}_b\)). We use these latent embeddings to calculate the edge weight between vertices in the graph. 
The edge weight between brands \(b_i\) and \(b_j\) is then defined as $E_{ij}=$  \(\textbf{v}_{b_i} \cdot \textbf{v}_{b_j} \) where $E_{ij} \in \mathbf{E}$.

\subsubsection{Sizing Recommendation Graph}

Further, we construct a multi-edge connected directed graph in which each node would have the gender, article type and brand information for a SKU. Two such nodes would be connected with 5 edges. Each edge would have a key "-1", "-0.5", "0", "0.5" and "1" and the value would represent the number of times two nodes were bought together having the size difference as the key. This has been shown in the Figure \ref{fig:size-graph} where two casual shoe nodes : Adidas and Clarks are connected using five edges. Edge '-1' represents how many times any shoe size of Clarks was bought one size lower than the same for Adidas. The same goes for other edges also. 
The Brand size graph helps to learn the co-relation between different sizes of various brands apart from recording the purchase data in a compact data structure.
\\
Formally, we construct a directed multi-edge size graph $G_s = (\mathbf{V},\mathbf{E})$ such that $\mathbf{V} = \mathbf{B}$ and $\exists$ \ 5 \ edges \ $\{e_{-1}, e_{-0.5}, e_{0}, e_{0.5}, e_{1}\}$ between every pair $(b_u, b_v)$ of vertices such that:

\begin{flalign*}
& e_{-1} : co-purchase \ frequency \ such \ that \ size(b_u) - size(b_v) = -1 & \\
& e_{-0.5} : co-purchase\  frequency \ such \ that \ size(b_u) - size(b_v) = -0.5 & \\
& e_{0} : co-purchase \ frequency \ such \ that \ size(b_u) - size(b_v) = 0 & \\
& e_{0.5} : co-purchase \ frequency \ such \ that \ size(b_u) - size(b_v) = 0.5 & \\
& e_{1} : co-purchase \ frequency \ such \ that \ size(b_u) - size(b_v) = 1 &
\end{flalign*}

\subsubsection{Probability Computation}
 The next and final section in recommending footwear product sizes to users is the Probability Computation. As shown in Figure \ref{fig:arch}, we have discussed about the creation of the Brand-Brand Graph $G_{b}$ and Size Recommendation Graph $G_{s}$ in the previous two sections. Let $u$ be the brand for which the user inputs the explicit shoe preference size and brand and $v$ be the unknown brand for which size needs to be predicted.
 \\
 Consider a subgraph \(G' \subseteq G_s \) such that \( V = \{u, v\} \) and \( E = \{e_{-1}, e_{-0.5}, e_{0}, e_{0.5}, e_{1} \} \). If there exists a direct connection $E$ between the vertices $V$, we compute total edge weight $e_{t}$ as :
 \\
 \\
 \begin{equation}
      e_{t} = \sum e_i \ \forall i \in \{ -1, -0.5, 0, 0.5, 1\} 
 \end{equation}
 
If $e_{t}$ exceeds a threshold \(\alpha\), we would regard it as a strong edge. \(\alpha\) is the measure of the strength of $e_{t}$ and is a tune-able hyperparameter. We then apply a softmax function over all the edge weights  \( w(e_{i}) \ where \  i \in  \{ -1, -0.5, 0, 0.5, 1\} \) :
\\
\begin{equation}
w'(e_{i}) = softmax(w(e_{i})) \ \forall i \in \{-1, -0.5, 0, 0.5, 1 \}
\end{equation}

If the \(max(w'(e_{i})) \ where \ i \in \{-1, -0.5, 0, 0.5, 1 \} \) exceeds a softmax threshold \(\lambda \), we choose the edge $e_{i}$ corresponding to the maximum edge weight as the path and make the size recommendation for $v$ as :

 \begin{equation}
     s(v) = s(u) - i  \ where \ i \in \{ -1, -0.5, 0, 0.5, 1\} 
     \label{eq:size-diff}
 \end{equation}
 
\begin{figure}
\includegraphics[height=2.0in, width=3.5in]{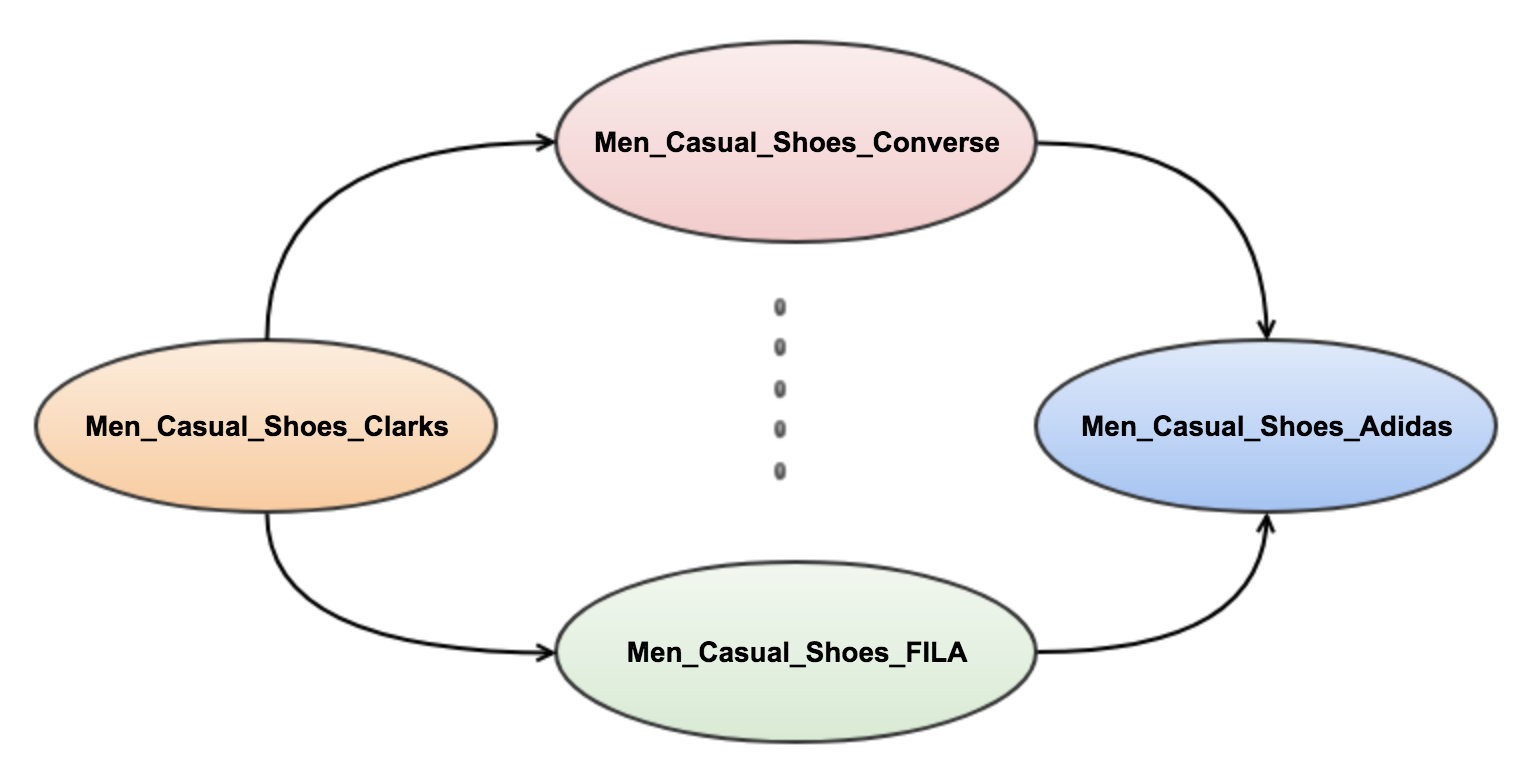}
\caption{Marginalized Probability}
\label{fig:size-marg}
\end{figure}

 Alternatively, if $u$ and $v$ are not directly connected, or if a strong edge is not discovered or if \( max(w'(e_{i})) < \lambda \) , we compute the probability of the edge weights by the rule of marginalization. We extract all the paths \( P = \{ u, z, v\} \) between $u$ and $v$ of path length exactly 2 such that  \(\textbf{Z}\) forms the set of all the intermediate vertices $z$ along the path of $u$ and $v$ as shown in the Figure \ref{fig:size-marg}. Then, we compute the recommended size for $v$ as :
 
 \begin{equation}
         P(s(v) \mid s(u)) = \sum_{z\in Z}P(s(z) \mid s(u)) * sim(u,z) * P(s(v) \mid s(z)) * sim(v,z)
 \end{equation}

 where $s(k)$ is the size for brand $k$. $sim(a,b)$ between two brands $a$ and $b$ is the similarity score and can be computed from $G_{b}$.
 
 For a better understanding, consider $u$ = 'Clarks' and $v$ = 'Adidas' in Figure \ref{fig:size-marg}. There are several brands in between $u$ and $v$ which connect them indirectly. For the sake of simplicity, consider \textbf{$Z$} consisting of only two brands for now, i.e., \textbf{Z} = \{Converse, FILA\}. Our aim is to predict the size for a casual shoe of brand "Adidas" given we have the explicit shoe preference information of brand "Clarks". Each node would be connected with five edges \( \{e_{-1}, e_{-0.5}, e_{0}, e_{0.5}, e_{1} \} \). We compute the marginalized probabilities for each of the paths from $u$ to $v$ by multiplying the edge weights and the similarity scores. For instance, if we want to calculate the marginalized probability of \(s(u) - s(v) = 1 \), we would consider the paths : \( \{ e_{u,z,1},e_{u,z,0} \}, \{ e_{u,z,0},e_{u,z,1}\} \ and \ \{ e_{u,z,0.5}, e_{u,z,0.5}\} \) where \(e_{m,n,i}\) is the edge weight between two brands $m$ and $n$ with the edge label $i$. The edge weights for each of the paths would be multiplied by the similarity score of the brands that edge is connecting and would be summed over all the possible paths and connecting brands (in this case Converse and FILA) to give a probability score for \(s(u) - s(v) = 1 \). The probability scores for all other size differences   \(S_{d} = s(u) - s(v) \ where \ S_{d} \in \{ -2, -1.5, -1, -0.5, 0, 0.5, 1, 1.5, 2 \} \) would also be calculated in a similar way by traversing the graph paths and combining the individual edge weights. \(s(v)\) would be calculated by the following equation where $i$ has the maximum log normalized score in \(S_{d}\):
 
  \begin{equation}
     s(v) = s(u) - i  \ where \ i \in S_{d}
     \label{eq:size-diff}
 \end{equation}
 
In the WBSR approach, we have implemented the rule of marginalization for probability computation for path length = 2. This is because, as observed in the train set 99\% of the brands which were not directly connected from every other brand are reachable through one-hop. 

\section{Results and Analysis}

This section demonstrates the results and analysis of the experiments we conducted. Firstly, we discuss about the dataset we have used, thereafter we proceed to discuss the quantitative and qualitative results for both the Skip-gram based recommendation and WBSR.
\subsection{Dataset}
For both the approaches mentioned in Section 2, we split our dataset into non-overlapping training, testing and cross-validation sets of model parameters. Both the approaches build separate models for each gender-article type (Men Casual Shoes, Men Sports Shoes, Women Heels etc). The order numbers and number of unique brands for train and test sets are tabulated in Table \ref{table:dataset}.

\begin{table}
\caption{Dataset statistics}
\begin{center}
\begin{tabular}{ |p{1cm}|p{2cm}|p{1.5cm}|p{1.5cm}|p{1.5cm}| }
 \hline
 Gender & Article Type & No. of orders (Train) & No. of orders (Test) & No. of unique brands \\
 \hline
 Men & Casual Shoes & 3.2M & 475K & 295 \\
 \hline
 Men & Sports Shoes & 1.1M & 33.6K & 68 \\
 \hline
 Men & Formal Shoes & 697K & 25.7K & 170 \\
 \hline
 Men & Flip Flops & 1M & 19.2K & 118 \\
 \hline
 Men & Sandals & 359K & 11.2K & 181 \\
 \hline
 Women & Heels & 826K & 31.9K & 193 \\
 \hline
 Women & Sports Shoes & 205.3K & 7.1K & 47 \\
 \hline
\end{tabular}
\label{table:dataset}
\end{center}
\end{table}
\subsection{Quantitative Results}

\begin{table}
\caption{Accuracy scores for various article types}
\begin{center}
\begin{tabular}{ |p{1cm}|p{2cm}|p{1.5cm}|p{1.5cm}| }
 \hline
 Gender & Article Type & Skip Gram Accuracy \% & WBSR Accuracy \% \\
 \hline
 Men & Casual Shoes & 80.85 & 79.25\\
 \hline
 Men & Sports Shoes & 83.8 & 87.8 \\
 \hline
 Men & Formal Shoes & 87.24 & 87.82 \\
 \hline
 Men & Flip Flops & 81.18 & 84.48 \\
 \hline
 Men & Sandals & 84.13 & 87.5 \\
 \hline
 Women & Heels & 73.69 & 72.69 \\
 \hline
 Women & Sports Shoes & 86.62 & 89.01 \\
 \hline
\end{tabular}
\label{table:accuracy}
\end{center}
\end{table}

Table \ref{table:accuracy} shows the accuracy numbers for the Skip Gram based and WBSR approaches. We see that the WBSR approach performs better than the Skip Gram based approach for 5 article types: Men Sports Shoes, Men Formal Shoes, Men Flip Flops, Men Sandals and Women Sports Shoes. This is because the number of distinct brands for these article types are a lot as compared to Women Heels and Men Casual Shoes, which results in poor latent vector representations due to sparsity. Capturing brand-level information and strengthening the product size relationship by leveraging the brand-brand similarity score helps in providing better and confident recommendations. In the following sections, we would be detailing the analysis of each approach.

\subsection{Skip-gram approach analysis}

The skip gram based approach for making size recommendations as explained in Section 2.1 computes Word2vec vectors for each SKU at a size level. Table \ref{table:brand-size} tabulates some products in the first column, which when served as a query, result in the size recommendations for brands in the second column. There are certain challenges for this approach which we can explain through the examples mentioned in Table \ref{table:brand-size}. To start with, products on our shopping platform can be segregated into various price brands like premium, mass premium, bridge to luxury and luxury \cite{Stand} and it's often seen that customers buy from a particular price band \cite{Stand,lynn2011segmenting, bhanot2013study}. As a result, we don't have much co-purchase history of products which are bought across price brands. Due to this, we don't have enough training samples to train our Word2vec network which results in local clustering of product-sizes with respect to price-bands in the size and fit space. This can be illustrated from the example of Froskie UK9 (Men Formal Shoes) in Table \ref{table:brand-size} which has a cosine similarity of 0.77 with a Timberland  UK8 (Men Formal Shoes). The physical measurements of these two sizes match but due to a high disparity in the brands' price range, the score is very less. Another challenge observed is the less user purchase history for certain brands as they have a very less visibility on our platform. To back this, observe in Table \ref{table:brand-size} that for a Superdry UK8 (Men Sports Shoes), an Adidas UK10.5 (Sports Shoes) would be recommended. This is because Superdry Sports Shoes have a very less visibility and volume on our platform and there are not enough samples to learn the relation between the individual sizes. However, for brands within a price-band and with a high volume of orders' data, this approach performs very well and learns good representations between the sizes as seen in the case for Inc 5 EURO36 (Women Heels) in Table \ref{table:brand-size}.

\begin{table}
\caption{Brand-size similarity}
\begin{center}
\begin{tabular}{ |p{3cm}|p{5cm}| } 
 \hline
 Brand\_Size & Similar Brand\_Size \\
 \hline
 Superdry\_UK8 (Men Sports Shoes) & Asics\_Tiger\_UK8 (0.97), Saucony\_UK8 (0.95), Adidas\_UK10.5 (0.95)  \\
 \hline
 Aeropostale\_UK11 (Men Casual Shoes) & Allen\_Solly\_UK11 (0.92), Nike\_UK10.5 (0.82), Converse\_5 (0.77) \\
 \hline
 Froskie\_UK9 (Men Formal Shoes) & Kenneth\_Cole\_UK9.5 (0.91), Aldo\_UK9 (0.89), Timberland\_UK8 (0.77) \\
 \hline
 Inc\_5\_EURO36 (Women Heels) & Rocia\_EURO36 (0.9), Mochi\_EURO35 (0.78), Crocs\_EURO33.5 (0.55) \\
 \hline
\end{tabular}
\label{table:brand-size}
\end{center}
\end{table}
\subsection{WBSR approach analysis}

The analysis for the second approach (WBSR) contains results for the brand-brand similarity index, quantitative results for the attributes of the Size Recommendation Graph and Accuracy numbers and plot for hyperparameters.

\subsubsection{Brand Brand graph}
As explained in the Section 2.2.1, the brand-brand graph computes the similarity scores between any two brands basis the user sessions' data. This score would help us to learn the relevancy of a brand's size in making the size recommendation for a second brand. The similarity scores are used during the probability computation step and help to strengthen the confidence in making correct size recommendations. Table \ref{table:brand-sim} enlists some of the brands in column 1 and the similarity scores for different brands in column 2.

\begin{table}
\caption{Similar Brands}
\begin{center}
\begin{tabular}{ |p{2cm}|p{6cm}| } 
 \hline
 Brand & Similar Brands \\
 \hline
 FOREVER 21 (Women Heels) & Carlton London (0.35), Steve Madden (0.34), Dorothy Perkins (0.25)  \\
 \hline
 Nike (Men Sports Shoes) & Adidas  (0.65), Puma (0.53), Reebok (0.50)  \\
 \hline
 Geox (Men Formal Shoes) & Johnston \& Murphy (0.44), Aldo (0.34), Kenneth Cole (0.30)  \\
 \hline
\end{tabular}
\label{table:brand-sim}
\end{center}
\end{table}
\subsubsection{Size Recommendation graph structure}
Table \ref{table:sparsity} tabulates the number of vertices, number of edges and the sparsity index for the probabilistic graph models $G_s$ built at a gender-article type level as explained in Section 2.2.2. The total number of vertices symbolize the nodes of the graph which contain the gender, article type and brand information for the SKUs. Remember that, two nodes can be connected via at most 5 non-zero weighted edges. Each has been counted separately while calculating the number of edges in Table \ref{table:sparsity} as this gives us a true representation of the user co-purchase history. Since sparsity is a measure of how completely-connected a graph is, we measure the connections between the nodes of $G_s$ only once. As we can see from the Table \ref{table:sparsity}, $G_s$ for men footwear are more sparse than women footwear. This can be attributed to the higher volume of orders and distinct brands available for men footwear.

\begin{table}
\caption{Size Recommendation Graph Structure}
\begin{center}
\begin{tabular}{ |c|c|c|c|c| }
 \hline
 Gender & Article Type & No. of vertices & No. of edges & Sparsity \\
 \hline
 Men & Casual Shoes & 271 & 31508 & 0.63  \\
 \hline
 Men & Sports Shoes & 62 & 1933 & 0.57 \\
 \hline
 Men & Formal Shoes & 157 & 8029 & 0.78 \\
 \hline
 Men & Flip Flops & 109 & 5437 & 0.61 \\
 \hline
 Men & Sandals & 167 & 7886 & 0.72 \\
 \hline
 Women & Heels & 179 & 17072 & 0.56 \\
 \hline
 Women & Sports Shoes & 39 & 703 & 0.66 \\
 \hline
\end{tabular}
\label{table:sparsity}
\end{center}
\end{table}
\subsubsection{Hyperparameters}

\begin{figure}
\includegraphics[height=2.0in, width=2.8in]{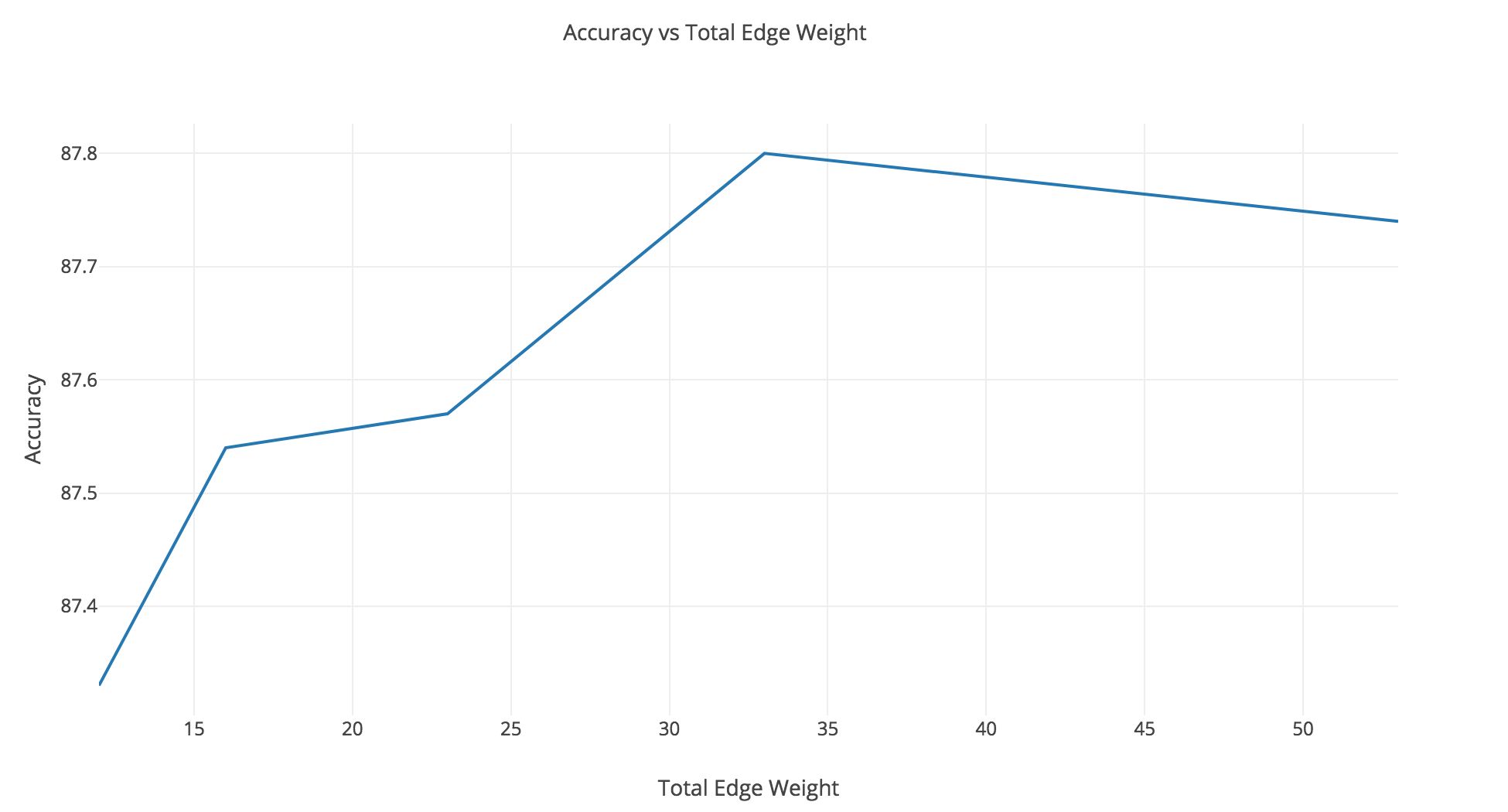}
\caption{Comparison of accuracy numbers for various total edge weight threshold "\( \alpha \)" for Men Sports Shoes}
\label{fig:edge-wt}
\end{figure}

In Section 2.2.2, we mentioned about the two hyperparameters we used while making size recommendations. The hyperparameter $\alpha$ signifies the strength of the edge weight while $\lambda$ signifies the softmax threshold. On analyzing the accuracy scores for different iterations of  $\lambda$ on the validation set, we observe that $\lambda$ = 0.7 gives us the highest accuracy numbers. Hence, we fix $\lambda$ at 0.7 and conduct an experiment over the validation set for achieving the most optimal value for $\alpha$. We ran the experiment on the validation set of Men Sports Shoes, wherein we chose the $60^{th}, 65^{th}, 70^{th}, 75^{th} and \ 80^{th}$ percentile values of total edge weights for $\alpha$. Figure \ref{fig:edge-wt} shows the accuracy numbers for different $\alpha$ values. We see that the maximum accuracy is achieved at $75^{th}$ percentile of total edge weight values. On similar lines, $\alpha$ for other article types can also be computed. 

\section{Conclusion}
In this paper, we have described a footwear size recommendation system which leverages the shoe preference questionnaire filled by the users to make size recommendations in the online fashion industry. This can be used to solve the cold-start problem wherein we don't have any order history for the user. We have primarily discussed two methods and have recorded the results for the same. The first method is skip-gram based which uses latent vector representations of the SKUs to make recommendations. We further discussed the challenges associated with this approach, namely high sparsity and less order purchase history which result in local clustering in the embedding space. To deal with this, we discuss the Weighted Brand Similarity Recommendation (WBSR) system which is a probabilistic graphical approach and works at a brand level to reduce sparsity. We reduced the sparsity from 0.8 in the skip-gram based approach to 0.64 on average.  We have also discussed about the method of incorporating the brand-brand similarity index in WBSR to give confident and valid size recommendations. We also discuss the method to apply the rule of marginalization in the probability computation step. This work can be extended for the scope of various other types of recommendations, like for size recommendation of Apparels. As a future work, we would like to extend our work for making subcategory level recommendations and also incorporate the categorical features for footwear while keeping the sparsity in check. 


  

\bibliographystyle{IEEEtran}
\bibliography{sample-bibliography}

\begin{thebibliography}{10}
\providecommand{\url}[1]{#1}
\csname url@samestyle\endcsname
\providecommand{\newblock}{\relax}
\providecommand{\bibinfo}[2]{#2}
\providecommand{\BIBentrySTDinterwordspacing}{\spaceskip=0pt\relax}
\providecommand{\BIBentryALTinterwordstretchfactor}{4}
\providecommand{\BIBentryALTinterwordspacing}{\spaceskip=\fontdimen2\font plus
\BIBentryALTinterwordstretchfactor\fontdimen3\font minus
  \fontdimen4\font\relax}
\providecommand{\BIBforeignlanguage}[2]{{%
\expandafter\ifx\csname l@#1\endcsname\relax
\typeout{** WARNING: IEEEtran.bst: No hyphenation pattern has been}%
\typeout{** loaded for the language `#1'. Using the pattern for}%
\typeout{** the default language instead.}%
\else
\language=\csname l@#1\endcsname
\fi
#2}}
\providecommand{\BIBdecl}{\relax}
\BIBdecl

\bibitem{abdullasize}
G.~M. Abdulla and S.~Borar, ``Size recommendation system for fashion
  e-commerce.''

\bibitem{internetRetailer}
\BIBentryALTinterwordspacing
T.~Mapel, ``Ringing in new year with rush of online returns,'' 2015. [Online].
  Available:
  \url{https://www.digitalcommerce360.com/2015/12/31/ringing-new-year-rush-online-returns/}
\BIBentrySTDinterwordspacing

\bibitem{hsu2009data}
C.-H. Hsu, ``Data mining to improve industrial standards and enhance production
  and marketing: An empirical study in apparel industry,'' \emph{Expert Systems
  with Applications}, vol.~36, no.~3, pp. 4185--4191, 2009.

\bibitem{hu2015collaborative}
Y.~Hu, X.~Yi, and L.~S. Davis, ``Collaborative fashion recommendation: a
  functional tensor factorization approach,'' in \emph{Proceedings of the 23rd
  ACM international conference on Multimedia}.\hskip 1em plus 0.5em minus
  0.4em\relax ACM, 2015, pp. 129--138.

\bibitem{adomavicius2005toward}
G.~Adomavicius and A.~Tuzhilin, ``Toward the next generation of recommender
  systems: A survey of the state-of-the-art and possible extensions,''
  \emph{IEEE transactions on knowledge and data engineering}, vol.~17, no.~6,
  pp. 734--749, 2005.

\bibitem{aroradecoding}
S.~Arora and D.~Warrier, ``Decoding fashion contexts using word embeddings.''

\bibitem{bracher2016fashion}
C.~Bracher, S.~Heinz, and R.~Vollgraf, ``Fashion dna: Merging content and sales
  data for recommendation and article mapping,'' \emph{arXiv preprint
  arXiv:1609.02489}, 2016.

\bibitem{sembium2017recommending}
V.~Sembium, R.~Rastogi, A.~Saroop, and S.~Merugu, ``Recommending product sizes
  to customers,'' in \emph{Proceedings of the Eleventh ACM Conference on
  Recommender Systems}.\hskip 1em plus 0.5em minus 0.4em\relax ACM, 2017, pp.
  243--250.

\bibitem{mikolov2013efficient}
T.~Mikolov, K.~Chen, G.~Corrado, and J.~Dean, ``Efficient estimation of word
  representations in vector space,'' \emph{arXiv preprint arXiv:1301.3781},
  2013.

\bibitem{aroradeciphering}
S.~Arora, A.~Madvariya, D.~Alok, and S.~Borar, ``Deciphering fashion
  sensibility using community detection.''

\bibitem{lee2001algorithms}
D.~D. Lee and H.~S. Seung, ``Algorithms for non-negative matrix
  factorization,'' in \emph{Advances in neural information processing systems},
  2001, pp. 556--562.

\bibitem{Stand}
E.~Corbellini and S.~Saviolo, ``Management of fashion and luxury companies,''
  \url{https://www.coursera.org/learn/mafash/lecture/PPEm1/fashion-market-segmentation},
  accessed 11/02/17.

\bibitem{lynn2011segmenting}
M.~Lynn, ``Segmenting and targeting your market: Strategies and limitations,''
  2011.

\bibitem{bhanot2013study}
S.~Bhanot and S.~R. Srinivasan, ``A study of the indian apparel market and the
  consumer purchase behaviour of apparel among management students in mumbai
  and navi mumbai,'' \emph{Navi Mumbai}, 2013.

\end{thebibliography}

\end{document}